\documentclass[aps,prc,twocolumn,superscriptaddress,showpacs]{revtex4}
\usepackage{graphicx}
\newcommand{\dg}{$^{\circ}$}
\newcommand{\beq}{\begin{equation}}
\newcommand{\eeq}{\end{equation}}

\begin{document}

%Title of paper
\title{Fine structure of the isovector giant dipole resonance in $^{208}$Pb: Characteristic scales and level densities}

\newcommand{\NIRS}{National Institute of Radiological Sciences, Chiba 263-8555, Japan}
\newcommand{\RCNP}{Research Center for Nuclear Physics, Osaka University, Ibaraki, Osaka 567-0047, Japan}
\newcommand{\WMI}{Department of Physics, Western Michigan University, Kalamazoo, MI 49008-5252, USA}
\newcommand{\MSU}{National Superconducting Cyclotron Laboratory, Michigan State University, MI 48824, USA}
\newcommand{\TUDarmstadt}{Institut f\"ur Kernphysik, Technische Universit\"{a}t Darmstadt, D-64289 Darmstadt, Germany}
\newcommand{\UCT}{Department of Physics, University of Cape Town, Rondebosch 7700, South Africa}

\author{I.~Poltoratska}\affiliation{\TUDarmstadt}
\author{R.~W.~Fearick}\affiliation{\UCT}
\author{A.~M.~Krumbholz}\affiliation{\TUDarmstadt}
\author{E.~Litvinova}\affiliation{\WMI}\affiliation{\MSU}
\author{H.~Matsubara}\affiliation{\RCNP}\affiliation{\NIRS}
\author{P.~von~Neumann-Cosel}\email{vnc@ikp.tu-darmstadt.de}\affiliation{\TUDarmstadt}
\author{V.~Yu.~Ponomarev}\affiliation{\TUDarmstadt}
\author{A.~Richter}\affiliation{\TUDarmstadt}
\author{A.~Tamii}\affiliation{\RCNP}

\date{\today}

\begin{abstract}
\begin{description}
\item[Background:] 
The electric isovector giant dipole resonance (IVGDR) in $^{208}$Pb has been measured with high energy resolution with the $(p,p')$ reaction under extreme forward angles [A. Tamii {\it et al.}, Phys. Rev. Lett. {\bf 107}, 062502 (2011)] and shows considerable fine structure.
\item[Purpose:] 
The aim of the present work is to extract scales characterizing the observed fine structure and to relate them  to dominant decay mechanisms of giant resonances. Furthermore, the level density of $J^\pi = 1^-$ states is determined in the energy region of the IVGDR. 
\item[Methods:] 
Characteristic scales are extracted from the spectra with a wavelet analysis based on continuous wavelet transforms.
Comparison with corresponding analyses of $B(E1)$ strength distributions from microscopic model calculations in the framework of the quasiparticle phonon model and the relativistic random phase approximation allow to identify giant resonance decay mechanisms responsible for the fine structure. 
The level density of $1^-$ states is related to local fluctuations of the cross sections in the energy region of the IVGDR, where contributions from states with other spin-parities can be neglected.
The magnitude of the fluctuations is determined by the autocorrelation function.  
\item[Results:] 
Scales in the fine structure of the IVGDR in $^{208}$Pb are found at 80, 130, 220, 430, 640, 960 keV, and at 1.75 MeV.
The values of the most prominent scales can be reasonably well reproduced by the microscopic calculations although they generally yield a smaller number of scales..
The inclusion of complex configurations in the calculations changes the E1 strength distributions but the impact on the wavelet power spectra and characteristic scales is limited.
The level density of $1^-$ states is extracted in the excitation energy range $9 -12.5$ MeV and compared to a variety of phenomenological and microscopic models. 
\item[Conclusions:]
In both models the major scales are already present at the one-particle one-hole  level indicating Landau damping as a dominant mechanism responsible for the fine structure of the IVGDR in contrast to the isoscalar giant quadrupole resonance, where fine structure arises from the coupling to low-lying surface vibrations.  
The back-shifted Fermi gas model  parameterization of Rauscher {\it et al.}, Phys. Rev. C \textbf{56}, 1613 (1997) describes the level-density data well, while other phenomeological and microscopic approaches fail to reproduce absolute values or the energy dependence or both. 
\end{description}
\end{abstract}

\

% insert suggested PACS numbers in braces on next line

\maketitle

\section{Introduction}

Giant resonances are elementary excitations of the nucleus and their understanding forms a cornerstone of microscopic nuclear theory.
They are classified according to their quantum numbers (angular momentum, parity, isospin). 
The isovector giant dipole resonance (IVGDR) has always played a central role because it was the first one observed experimentally and thus triggered many basic theoretical concepts for its description.
Gross properties of the IVGDR like energy centroid and strength in terms of exhaustion of the energy-weighted sum rule are well described in macroscopic as well as microscopic models \cite{har01}.
However, despite recent progress a systematic understanding of the decay width is still lacking. 

The giant resonance width $\Gamma$ is determined by the interplay of different mechanisms: fragmentation of the elementary one particle-one hole ($1p1h$) excitations (Landau damping $\Delta E$)), direct particle decay out of the continuum (escape width $\Gamma\!\uparrow$), and statistical particle decay due to coupling to two ($2p2h$) and many particle-many hole ($npnh$) states (spreading width  $\Gamma\!\downarrow$) 
\begin{equation}
\label{eq:width}
\Gamma = \Delta E + \Gamma\!\uparrow + \Gamma\!\downarrow.
\end{equation}
A powerful approach to investigate the role of the different components are coincidence experiments, where direct decay can be identified by the population of one-hole states in the daughter nucleus and the spreading width contribution can be estimated by comparison with statistical model calculations (see, e.g., Refs.~\cite{bol88,die94,hun03}).
Recently, an alternative method has been developed based on a quantitative analysis of the fine structure of giant resonances oberved in high-resolution inelastic scattering experiments.
For comparable energy resolution, the fine structure is independent of the exciting probe \cite{kam97}.
A case study of this method has been performed for the isoscalar giant qudrupole resonance (ISGQR) from medium-mass to heavy nuclei \cite{she04,she09}.
Different approaches for an extraction of energy scales characterizing the observed fine structure have been compared in Ref.~\cite{she08}. 
Wavelet analysis has been identified as a particularly promising type of analysis.

It could be shown that the fine structure of the ISGQR arises from the mixing of the $1p1h$ states with a particular class of $2p2h$ states, viz.\ those of $1p1h \otimes {\rm phonon}$ character.
The coupling to low-energy phonons has been predicted to be a main source of the spreading width \cite{ber83}.
Differences of the characteristic scales between the investigated nuclei could be traced back to their low-energy collective structure.
One exception is $^{40}$Ca - the lightest nucleus studied so far with the wavelet technique -, where a recent random-phase approximation (RPA) calculation employing a realistic nuclear interaction derived with the Unitary Correlation Operator Method (UCOM) found that the characteristic scales result from Landau damping \cite{usm11a} in contrast to a large variety of previous RPA results, where the ISGQR strength is always concentrated in a single state.   
  
\begin{figure}[tbh]
        \centering
 \includegraphics[width=8.6cm]{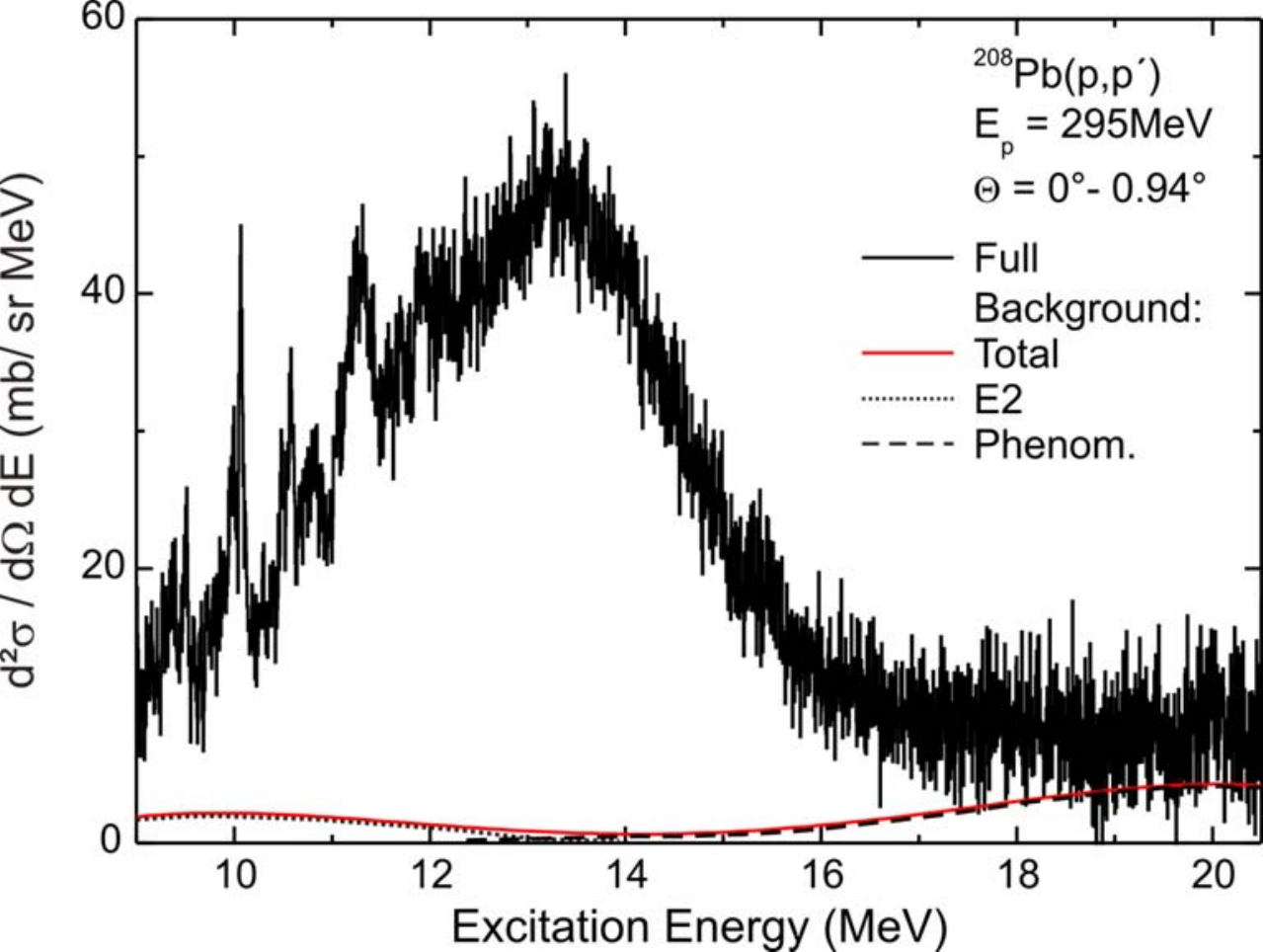}       
\caption{(Color online) Spectrum of the $^{208}$Pb$(p,p')$ reaction at $E_0 = 295$ MeV and $\Theta = 0^\circ - 0.94^\circ$.
The crosss sections are due to $E1$ excitations populated by relativistic Coulomb excitation.
Background from non-$E1$ excitations (full line) is determined by a MDA with contributions from excitation of $E2$ strength (dotted line) and a phenomenological component (dashed line) \cite{pol11}.}
                \label{fig:spectrum}
\end{figure}

 Here, we present a first application of the wavelet analysis to the IVGDR.
The doubly magic nucleus $^{208}$Pb is taken as a reference case because techniques to include states built on complex configurations beyond $1p1h$ states in RPA-type approaches are most advanced for closed-shell nuclei.
Experimentally, structure of the cross sections in the IVGDR energy region has already been observed in photonuclear reactions long ago, and its nature has been a subject of discussion \cite{sta82,bel82}.
Recently, proton scattering at energies of a few hundred MeV and under extreme forward angles including $0^\circ$ has been established as a new spectroscopic tool for the study of dipole strength with unprecedented resolution \cite{tam09,nev11}.
In these kinematics the cross sections of the $^{208}$Pb$(p,p')$ reaction are dominated by relativistic Coulomb excitation populating the IVGDR. 
Figure \ref{fig:spectrum} shows a spectrum at $E_0 = 295$ MeV and covering an angular range $\Theta = 0^\circ - 0.94^\circ$.
The full (red) line indicates the background from other contributions to the spectrum deduced by a multipole decomposition analysis (MDA) \cite{tam11,pol11}.
Different from the MDA in the low-energy region discussed in Ref.~\cite{pol12}, the main contributions are from excitation of the ISGQR (dotted line) and a phenomenological part (dashed line) including quasifree reactions and the tail of giant resonances centered at higher excitation energies.
In any case, the contributions under the IVGDR peak are small justifying the assumption that they do not influence the fluctuations visible in the data. 

The cross section fluctuations are particularly pronounced on the lower side of the IVGDR and are damped on the upper side.   
The magnitude of the fluctuations for a given experimental energy resolution is determined by the density of $1^-$ states.
If a single excitation mode dominates the cross sections -- as in the present case -- and there is a way to estimate the background in the spectra, one can deduce the level density of $1^-$ states  in the energy region of the giant resonance with a fluctuation analysis. 
Level densities are basic nuclear structure quantities and refined models such as shell-model Monte Carlo \cite{oze13}, quantum Monte Carlo \cite{hou09}, or a Hartree-Fock-Bogoliubov (HFB) plus combinatorial approach \cite{gor08} have been devoloped for their description.
Besides the MDA discussed above, an independent method to determine the background based on a wavelet analysis has been developed \cite{kal06}.
When either background subtraction procedure is applied to the $^{208}$Pb$(p,p')$ data, level densities of $1^-$ states in the energy region of the IVGDR can indeed be extracted and compared to a variety of phenomenological and microscopic models.
This experimental method to determine level densities is complementary to approaches based on compound nucleus $\gamma$-decay \cite{sie09} and particle emission \cite{voi07}, or thermal neutron capture \cite{mug06}.    

The article is structured as follows: 
In Sec.~\ref{subsec:cwt} the wavelet analysis technique of the experimental and theoretical spectra is introduced and applied in Sec.~\ref{subsec:cwt-208pb}.
Section \ref{sec:ld} deals with the extraction of level densities  with a description of the method in Sec.~\ref{subsec:fluctuationanalysis}, a discussion of methods for background subtraction in Sec.~\ref{subsec:DWTBackground}, and the application to the IVGDR in $^{208}$Pb in Sec.~\ref{subsec:leveldensityGDR}.  
The paper closes with conclusions (Sec.~\ref{sec:conclusions}).

\section{Characteristic scales from a wavelet analysis}
\label{sec:scales}

\subsection{Continuous wavelet transform (CWT)}
\label{subsec:cwt}

The wavelet transform is an established tool to analyze different types of signals hidden in fluctuating quantities, e.g.\ with time or energy. 
It is used in diverse areas, such as image processing or data compression \cite{lit_wavelet1,lit_wavelet2}, and also applied in meteorology \cite{lit_wavelet3}, astrophysics \cite{lit_wavelet4} or accelerator physics \cite{lit_wavelet5}.
The wavelet analysis can be regarded as an extension of the Fourier analysis which allows to conserve the correlation between the observable and its transform.

In the present case energy spectra of nuclear giant resonances are analyzed.
The coefficients of the wavelet transform are then defined as
\begin{equation}
   C\left( {\delta E,E_{x}} \right) = \int\limits_{ - \infty
   }^\infty  {\sigma\left( E \right)\Psi \left( {\delta E,E_{x},E}
   \right)dE}.
   \label{eq:cwt}
\end{equation}
They depend on two parameters, the scale $\delta$E stretching and compressing the wavelet $\Psi$(E), and the position E$_x$ shifting the wavelet in the spectrum $\sigma$(E). 
The variation of the variables can be carried out using continuous (CWT) or discrete (DWT) steps. 
The analysis of the fine structure of giant resonances is performed using CWT, where the fit procedure can be adjusted to the required precision. 
The application of DWT for an analysis of background contributions in the spectra is discussed in Sec.~\ref{sec:ld}. 
Applications of the CWT to high-resolution nuclear spectra of giant resonances are described in Refs.~\cite{she04,she09,usm11a,pet10}. 
Further details and a comparison with other techniques for the analysis of fine structure in nuclear giant resonances can be found in Ref.~\cite{she08}. 

The choice of the wavelet function plays an important role in the analysis. 
In order to achieve an optimum representation of the signal using wavelet transformation one has to select a function $\Psi$ which resembles the properties of the studied signal $\sigma$. 
In fact, the better the correspondence between the shape of $\Psi$ and the signal $\sigma$ is, the larger is the wavelet coefficient.
A maximum of the wavelet coefficients at certain value $\delta$E indicates a correlation in the signal at the given scale, often called characteristic scale.
The best resolution for nuclear spectra is obtained with the so-called Morlet wavelet (cf.\ Fig.~9 in Ref.~\cite{she08}) because the detector response is typically close to the Gaussian line shape and the Morlet wavelet is a product of Gaussian and cosine functions
\begin{equation}
\psi_{Morlet} (x) = \pi^{-1/4} e^{ikx} e^{-x^2/2}.
\label{eq:morlet}
\end{equation}
The results show little difference whether the complex Morlet function or only the real part is considered. 
Therefore, only the real Morlet function was used.  
 
\subsection{Application to the IVGDR in $^{208}$Pb} \label{sub:scales}
\label{subsec:cwt-208pb}

In the following, we apply a CWT analysis to the $^{208}$Pb$(p,p^\prime)$ excitation spectrum for scattering angles $\Theta$~=~0\dg~-~0.94\dg.
At these extreme forward angles $E1$ Coulomb excitation dominates the cross sections and nuclear transitions are suppressed with the exception of the isovector spin-flip $M1$ resonance. 
The excitation energy region below 9~MeV, where the spin-$M1$ mode is located and contributes significantly to the cross sections \cite{pol12,hey10}, is thus excluded.
In order to search for characteristic scales it is helpful to construct the power spectrum of the signal, i.e.\ the projection of the absolute values of the wavelet coefficients on the scale axis. 

\begin{figure}[b]
        \centering
 \includegraphics[width=8.6cm]{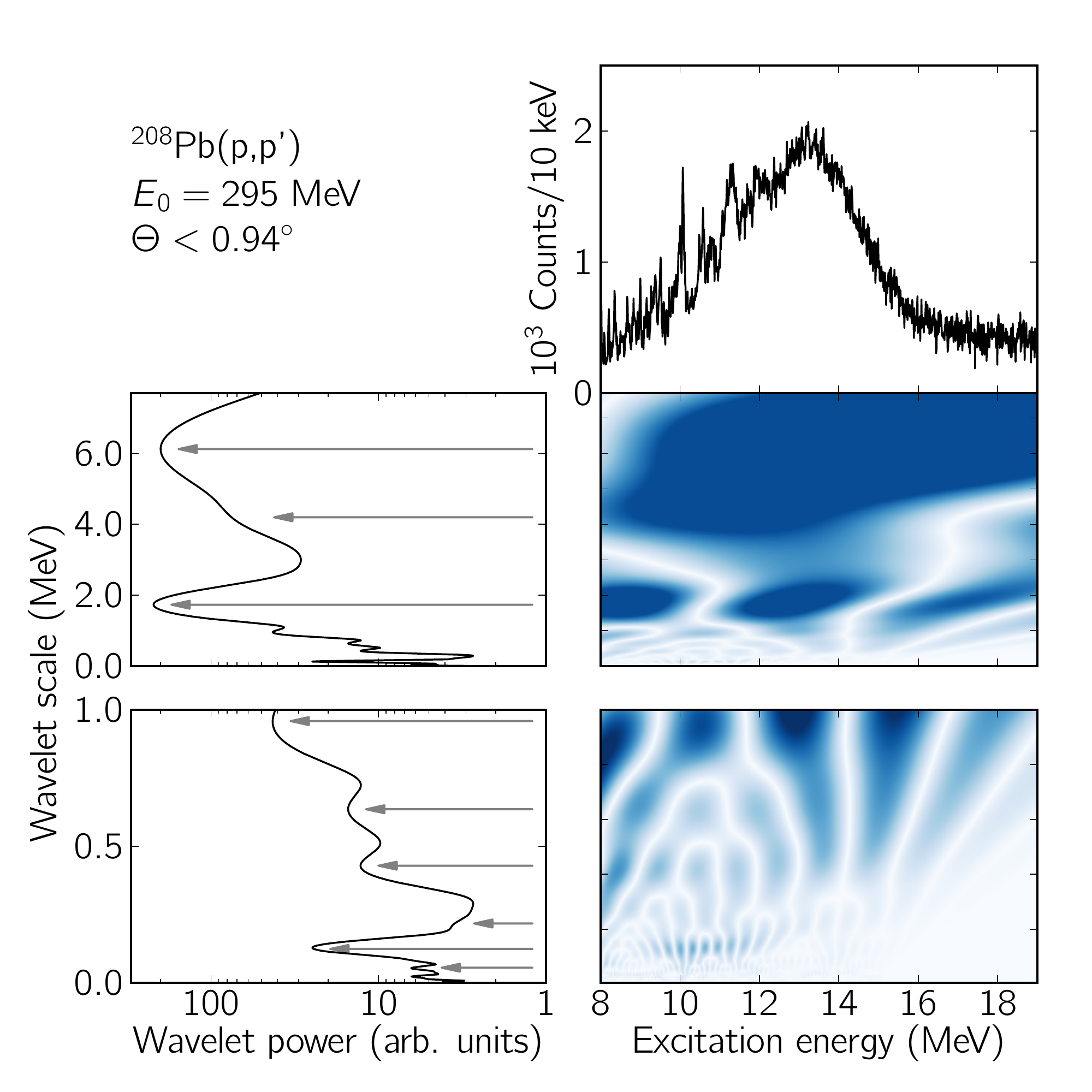}       
\caption{(Color online) CWT analysis of the excitation energy spectrum of the $^{208}$Pb($p,p^\prime$) reaction at E$_0 =295$~MeV and $\Theta_{lab} <$0.94\dg. Top-right: Spectrum of the reaction in the IVGDR region.
Middle: Absolute values of the wavelet coefficients (right) and power spectrum (left). Bottom: Enlarged picture for the region of scales below 1~MeV.
White color corresponds to smallest wavelet coefficients, while dark regions indicate the largest values. 
Arrows indicate the positions of characteristic scales.}
                \label{fig:cwt1}
\end{figure}
In Fig.~\ref{fig:cwt1} the excitation energy spectrum (upper right) and corresponding absolute values of the wavelet coefficients (middle and lower right) are plotted. 
White regions indicate the smallest values of the wavelet coefficients, while dark ones denote maxima, i.e.\ characteristic scales.
One identifies scale values where the absolute values of the wavelet coefficients show a local maximum, albeit with a characteristic minimum/maximum variation as a function of excitation energy induced by the oscillating wavelet function.
For a better recognition of such characteristic scales power spectra are plotted (middle- and lower-left). 
The power values are divided by the corresponding scale in order to remove a trivial increase with increasing scale \cite{liu07}. 
The middle panel shows the scale region up to 7 MeV, while the lower panel gives an enlarged view of the region below 1~MeV. 
Characteristic scales are clearly visible in the power spectra indicated by arrows.

The extracted scale values are converted to correspond to the full width at half maximum (FWHM) of a Lorentzian function, as described in Ref.~\cite{she08}.  
Characteristic scales are observed at  at 80, 130, 220, 430, 640, 960 keV, 1.75, 4, and 6 MeV.
Two scales are found below 100 keV, where the smallest scale at about 30~keV corresponds to the experimental energy resolution.
The strong scale at 130 keV is confined in energy to the region $9-12$ MeV where the most pronounced structure is seen in the IVGDR of $^{208}$Pb.
The other scales up to 1 MeV are related to a larger excitation energy region extending up to about 15 MeV, while the dominating scale at 1.75 MeV and the broad scale at large values appear over the whole resonance region.
A characteristic scale roughly corresponding to the width of the resonance of about 4~MeV is indicated as a shoulder of the bump peaking at about 6 MeV. 
It should be noted that at scale values of several MeV uncertainties due to the limited data range of about 10 MeV become dominant preventing a clear interpretation of the two largest scales.

\begin{figure}[tbh]
        \centering
 \includegraphics[width=8.6cm]{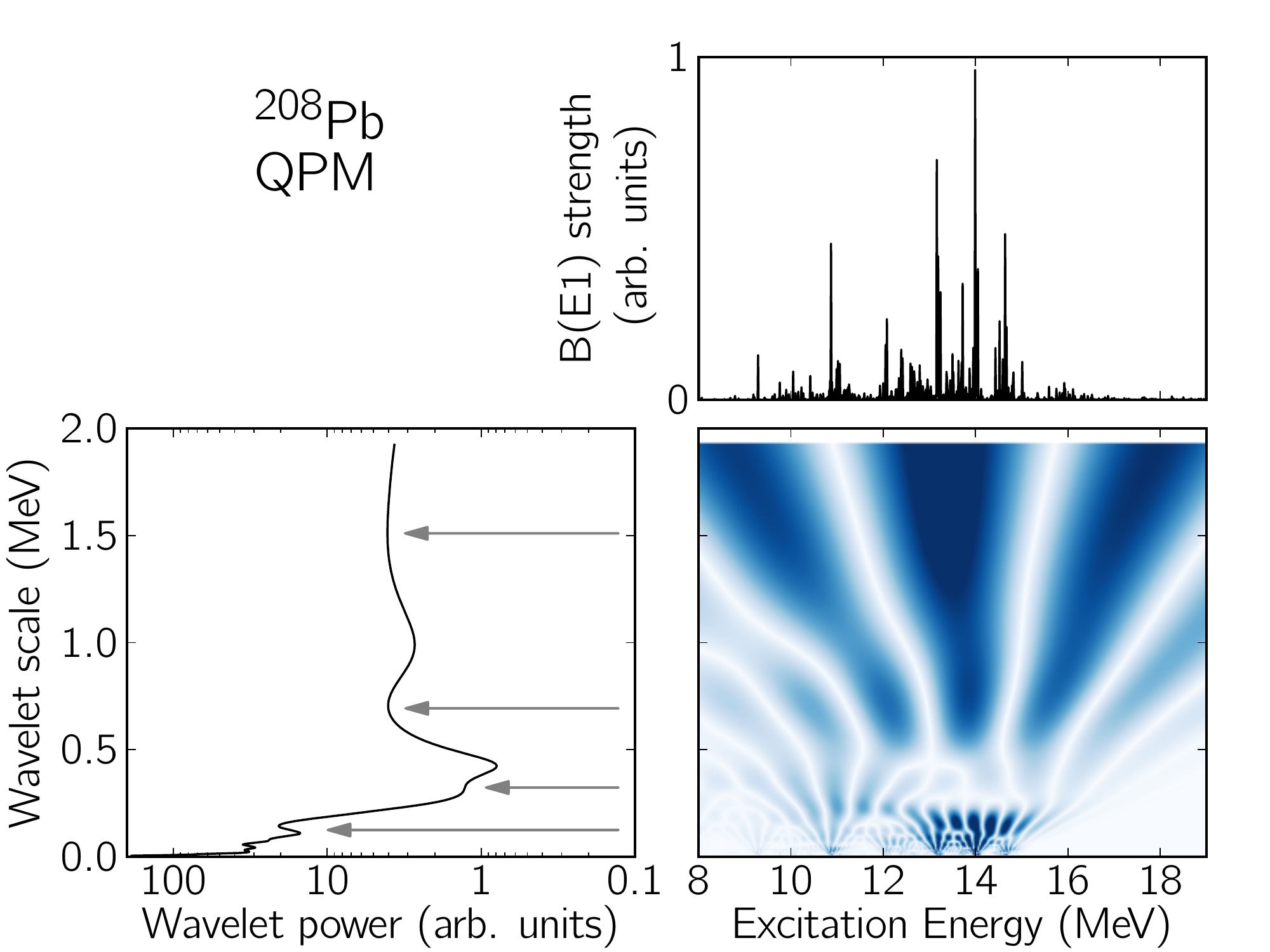}
        \caption{(Color online) CWT analysis of the IVGDR strength distribution from QPM calculations (upper r.h.s.) described in the text. 
White color corresponds to the smallest values of the wavelet coefficients, while dark blue shows the maximum. 
Arrows indicate the positions of prominent characteristic scales.}
                \label{fig:cwtqpm}
\end{figure}
\begin{figure}[tbh]
        \centering
 \includegraphics[width=8.6cm]{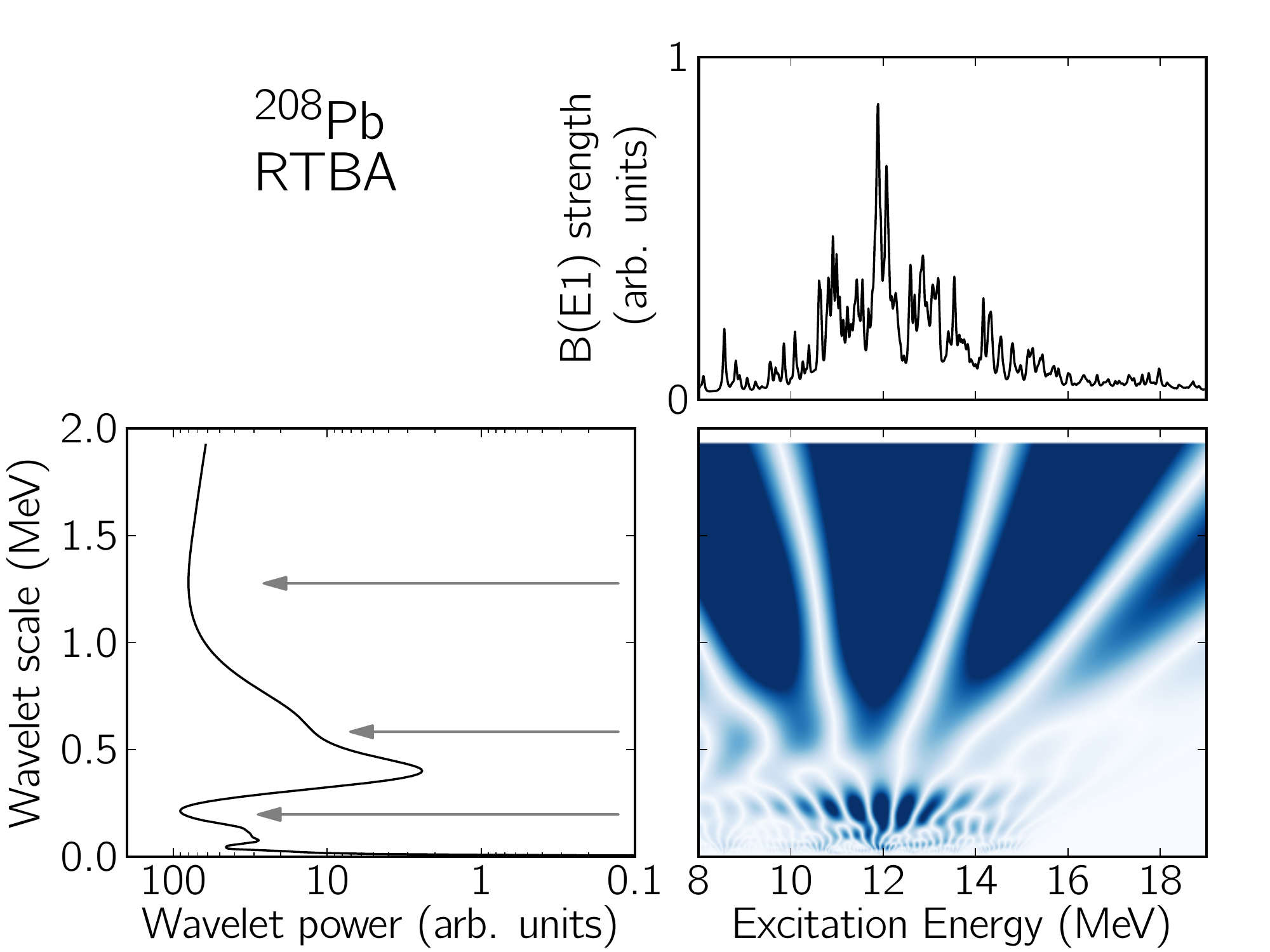}
        \caption{(Color online) CWT analysis of the GDR strength distribution from RTBA calculations (upper r.h.s.) described in the text. 
White color corresponds to the smallest values of the wavelet coefficients, while dark blue shows the maximum. 
Arrows indicate the positions of prominent characteristic scales.}
                \label{fig:cwtrtba}
\end{figure}
In order to understand the origin of the characteristic energy scales obtained from the experimental data one needs a comparison with theoretical calculations. 
Results of the CWT analysis for microscopic calculations of the electric dipole response in $^{208}$Pb with the quasiparticle phonon model (QPM)  and relativistic RPA (RRPA) are discussed.
Both models allow for the inclusion of complex configurations. 
Therefore, besides calculations on the  $1p1h$ level (called QPM 1-phonon and RRPA, respectively), also extensions including $2p2h$ states (called QPM and relativistic time blocking approximation (RTBA), respectively) are considered.
A general description of the QPM can be found in Ref.~\cite{sol92} and of the RTBA in Ref.~\cite{lit07}.
Details of the present QPM calculations are discussed in Refs.~\cite{tam11,pol12,rye02}. 
Results of the wavelet analysis of QPM and RTBA  E1 strength distributions are presented in Figs.~\ref{fig:cwtqpm} and \ref{fig:cwtrtba}, respectively.

The three most prominent scales (i.e., 140 keV, 720 keV, 1.55 MeV) observed in the QPM calculations are in fair agreement with values deduced from experiment. 
However, the relative power differs compared to experment with the most prominent scale in the QPM results at low energy while the equivalent of the strongest experimental scale at about 1.75 MeV is less pronounced.
Also, some experimentally observed scales do not show up in the calculation.
The picture obtained from the RTBA results is quite similar but the larger scale above 1 MeV is more and the 0.6 MeV scale less pronounced than in the QPM case. 
A summary of the extracted scales is given in Tab.~\ref{tab:scales} together with an analysis of the corresponding RPA results. 
The experimental scales at 4 and 6 MeV scales are not included in Tab.~\ref{tab:scales} for the reason discussed above.
\begin{table}[tbh!]
\caption{Characteristic scales of the GDR in $^{208}$Pb extracted from the wavelet analysis of the experimental data and from the QPM, RRPA and RTBA calculations described in the text.}
  \label{tab:scales}
  \begin{center}
\begin{tabular}{lccccccc}
\hline\hline
 & \multicolumn{7}{c}{Scales (keV)}  \\
\hline
Experiment & 80 & 130 &  220 & 430  & 640  &  960 & 1750  \\
QPM (1-phonon) &  100 & 160 & \multicolumn{2}{c}{340} & 720 & & 1550 \\
QPM  &  90 & 140 & \multicolumn{2}{c}{340} & 700 & & 1500 \\
RRPA & 90 & 140 & 210 & &580 & & 1050 \\
RTBA & 90 &  \multicolumn{2}{c}{220} & & 600 & & 1300 \\
  \hline\hline
    \end{tabular}
  \end{center}
\end{table}

\begin{figure}[t]
        \centering
\includegraphics[width=8.6cm]{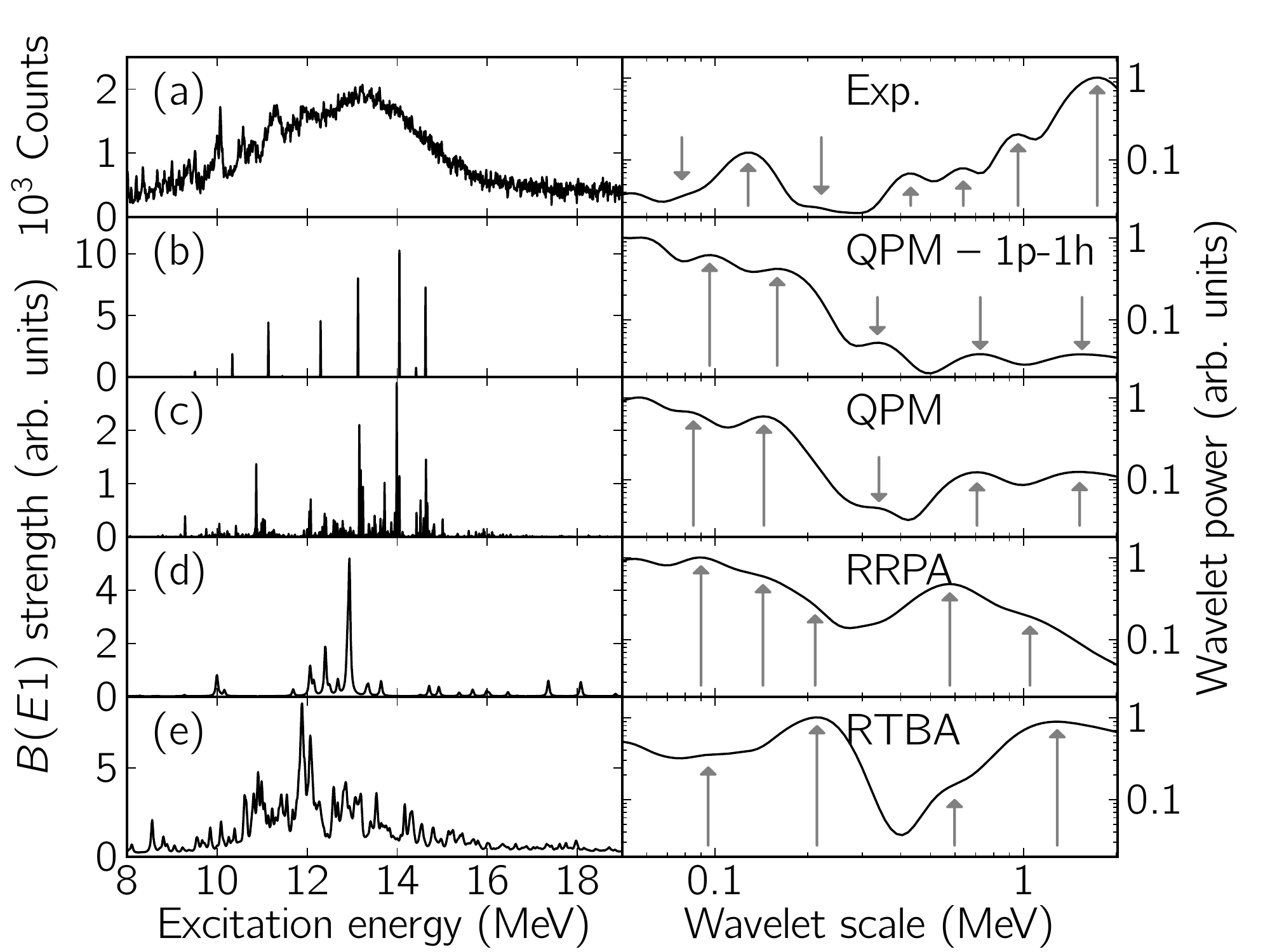}
        \caption{(a) Experimental spectrum of the $^{208}$Pb($p,p^\prime$) reaction of Fig.~\ref{fig:cwt1} in comparison with theoretical predictions of  the $B(E1)$ strength distribution in $^{208}$Pb (l.h.s.) and the resulting power spectra from a CWT analysis (r.h.s.). 
Theoretical results are shown for the QPM with 1-phonon (b) and (1+2)-phonon (c) model spaces, RRPA (d)  and RTBA (e). 
Characteristic scales are marked by arrows.}
                \label{fig:powerspectra}
\end{figure}

A comparison of the experimental cross sections at $0^\circ$ (l.h.s) and the power spectrum (r.h.s.) resulting from the CWT analysis with those of the model calculations for the $B(E1)$ strength distributions is shown in Fig.~\ref{fig:powerspectra}.
It should be noted that the experimental spectrum, (a), does not represent the $B(E1)$ strength but the Coulomb excitation cross section, which is modified by the excitation-energy dependent virtual photon number. 
Extraction of the $B(E1)$ distribution is possible (cf.~Refs.~\cite{tam11,pol12}). 
However, the need to disentangle the $E1$ cross section from other contributions can only be achieved for larger energy bins, where the information on the fine structure is partially lost.
Such a conversion of the experimental data to $B(E1)$ strength would lead to a slight shift ($< 5$\%) of the characteristic scales and an increase of relative power to higher excitation energies..   

As shown in Fig.~\ref{fig:powerspectra}(b), a QPM calculation on the RPA level results in a $B(E1)$ strength distribution dominated by 5 transitions distributed between 11 and 15 MeV with a centroid  energy of 13.25 MeV (defined as $m_1/m_0$, where $m_i$ denotes the $i^{\rm th}$ moment of the distribution).
The experimental centroid energy of 13.43 MeV is fairly well reproduced.
Inclusion of 2-phonon configurations, Fig.~\ref{fig:powerspectra}(c), leads to fragmentation but the dominant $1p1h$ transitions remain and the centroid energy is unaffected.
A similar comparison of RRPA, Fig.~\ref{fig:powerspectra}(d), and RTBA, Fig.~\ref{fig:powerspectra}(e), results shows somewhat larger differences of the distributions although the centroid energy is hardly changed (13.01 MeV for RRPA and 13.06 MeV for RTBA, respectively). 

Since there is no absolute scale, the corresponding CWT power spectra shown on the r.h.s.\ of Fig.~\ref{fig:powerspectra} are normalized relative to each other. 
They provide a qualitative measure for the ability of different models to describe fine structure and characteristic scales.
Overall, both models broadly reproduce the variation of power with scale value. 
A power peak at small scales of a 100 - 200 keV is followed by a minimum of power at a few hundred keV and another rise towards larger values.
The scale values of power maxima and minima are better reproduced by the QPM.
However, the relative ratio of maxima at smaller and larger scales is predicted to decrease in the QPM while experiment shows an increase.
In the RTBA the ratio is closer to the data.  
The region of scales in the figure is restricted to 2 MeV because the theoretical calculations show limited power at even larger scale values, in contrast to the experiment.
This finding may be related to the neglection of coupling to the continuum in the models.

The comparison of Figs.~\ref{fig:powerspectra}(b,c) and (d,e) allows to extract information on the damping mechanism responsible for the fine structure.
Clearly, the QPM results show structure already at the 1-phonon level.
While the appearance of scales $\geq 1$ MeV can be easily understood by the spacing of the five dominant transitions, the wavelet analysis  of the RPA result (b) also finds the characteristic scales with smaller values $<1$ MeV.  
The similarity between the power spectra and scales deduced from the QPM calculation for a one-phonon model space with those including two-phonon states suggests that the fragmentation of $1p1h$ transitions (i.e., Landau damping) is the most important mechanism leading to fine structure of the IVGDR in $^{208}$Pb. 
The coupling to complex configurations and, in particular, to low-lying collective vibrations identified as dominant mechanism in the ISGQR in heavy nuclei~\cite{she09} seems to play a minor role only.
While the relative weight changes, major scales are also found at about the same energies in the CWT analysis of the RRPA (d) and RTBA (e) results.
The observation of characteristic scales in the RRPA calculation again supports an interpretation of Landau damping as a main cause of the fine structure of the IVGDR in $^{208}$Pb.

\section{Level density of $J^\pi = 1^-$ states}
\label{sec:ld}

In this section the extraction of  the level density of $1^{-}$ states in $^{208}$Pb in the excitation energy region of the IVGDR  by means of a fluctuation analysis is described. 

\subsection{Fluctuation analysis}
\label{subsec:fluctuationanalysis}

To extract level densities from high-resolution spectra, a fluctuation analysis can be utilized. 
The method was originally proposed to analyze $\beta$-delayed particle emission spectra~\cite{jon76}, but later it was successfully adopted for the study of electron scattering data~\cite{mue83,kil87} and can be used in general for high-resolution spectra of nuclear reactions (see, e.g., Refs.~\cite{kal06,kal07,end97}). 
Detailed descriptions of the method can also be found in Refs.~\cite{han79,han90}.  
The main idea of the analysis is to take advantage of the autocorrelation function in order to obtain a measure of the cross-section fluctuations
with respect to a stationary mean value.

The method can be applied in an energy region where the mean level spacing $\langle D \rangle$ is smaller than the experimental energy resolution $\Delta E$. 
One has to distinguish between two possible cases: 
(i) $\left\langle \Gamma  \right\rangle \le \left\langle D \right\rangle$, i.e., the mean level width $\left\langle \Gamma \right\rangle$ is smaller than the average distance between levels and the fluctuations result from the density of  states and their incoherent overlap, and
(ii) $\left\langle \Gamma  \right\rangle > \left\langle D \right\rangle$, the so-called Ericson fluctuations~\cite{eri60}, which result from the coherent overlap of the states.
In principle, it is possible to utilize the method in the Ericson regime, but the statistics has to be very high because of the large number of open decay channels. 
Thus, in practice one is usually limited to the region $\left\langle \Gamma \right\rangle \le \left\langle D \right\rangle$.

The application of the fluctuation analysis is based on the following two assumptions:\\
(i) In an highly-excited nucleus, the probability for a given spacing between states with the same spin and parity is given by the Wigner distribution \cite{wig65}
    \begin{equation}
        P_{W}(s) \;\; = \;\; \frac{\pi s}{2} \; \exp
        \left( - \frac{\pi s^{2}}{4} \right),
        \label{eq:wigner}
\end{equation}
with $ s= D/\langle D \rangle$.
This distribution has a maximum close to the mean value and shows a suppression of small distances between neighboring levels. \\
(ii) The ground state decay widths or transition strengths obey a Porter-Thomas distribution \cite{por56}
    \begin{equation}
    \label{eq:Porter}
         P_{PT}(s) \;\; = \;\; \frac{1}{\sqrt{2 \pi s}}
         \; \exp \left( - \frac{s}{2} \right)
\end{equation}
with $s = \Gamma_{0} / \langle     \Gamma_{0} \rangle$. \\
These assumptions are adopted from random matrix theory (RMT)~\cite{wei09,mit10} and based on the observation that they provide a good description of nuclear excitations in the vicinity of the neutron separation energy~\cite{haq82}.

\begin{figure}[tbh!]
        \centering
\includegraphics[width=8.6cm]{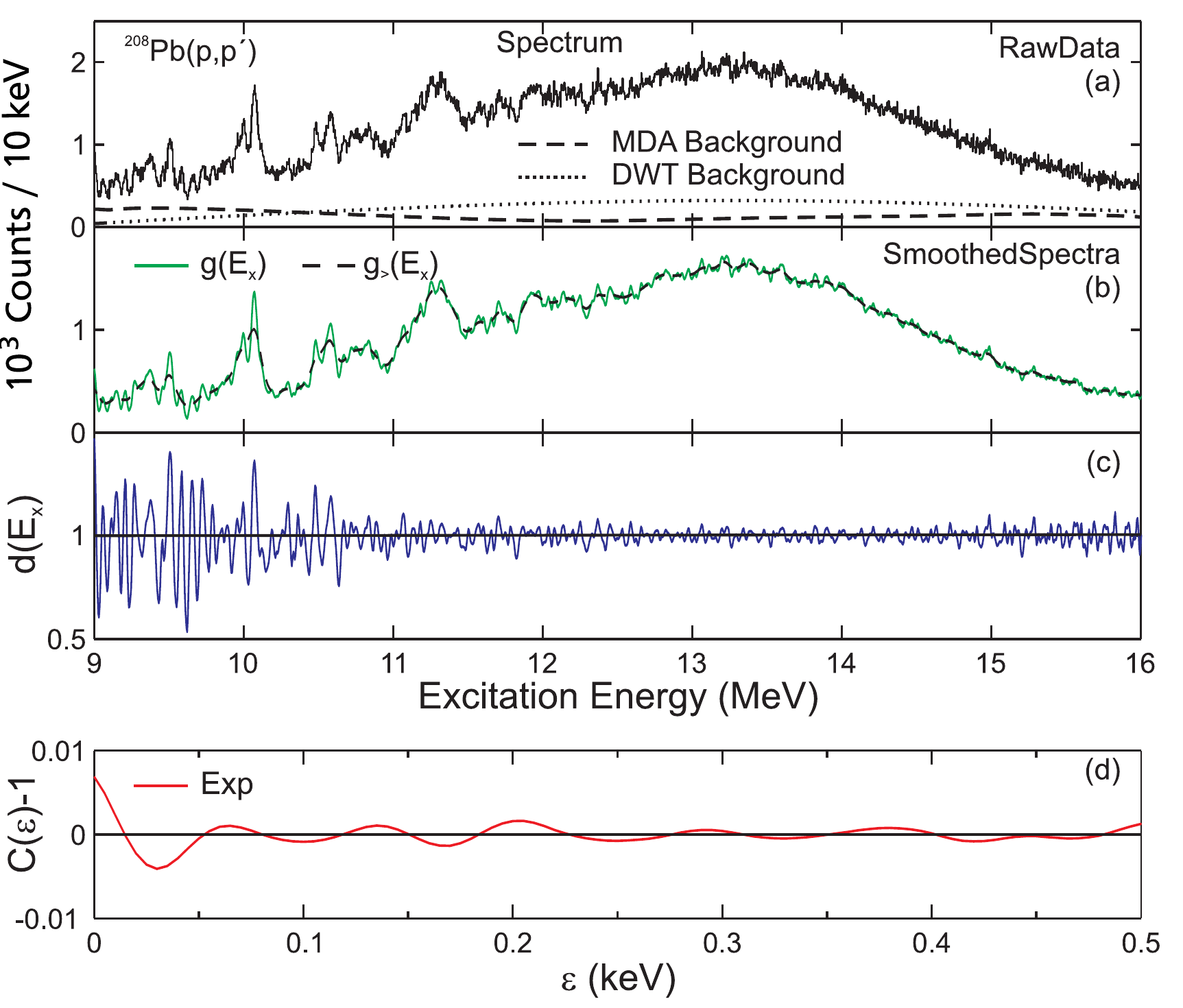}
        \caption{(Color online) (a) Spectrum of the $^{208}$Pb($p,p^\prime$) reaction at $E_0 = 295$~MeV and $\Theta = 0^\circ - 0.94^\circ$ and the background obtained from MDA (dashed line) and DWT (dotted line). (b) Background-subtracted smoothed spectra $g(E_x)$ and $g_>(E_x)$. (c) stationary spectrum $d(E_x)$.  (d) Experimental autocorrelation function.}
\label{fig:dwtprocedure}
\end{figure}

The procedure of the fluctuation analysis for the $^{208}$Pb($p,p^\prime$) scattering data at 0\dg~is schematically demonstrated in Fig.~\ref{fig:dwtprocedure}.
It can be divided in four main steps.
In order to obtain a spectrum containing only the information needed, one has to subtract any background not arising from excitations of the nuclear mode under investigation. Methods to determine this background are discussed in Sec.~\ref{subsec:DWTBackground}.

After background subtraction, the spectrum contains the information on the fluctuations in the spectrum of the IVGDR. 
In order to eliminate the fluctuation contributions arising from finite statistics, the spectrum is folded with a Gaussian function with the width $\sigma$ chosen to be smaller than the experimental energy resolution. 
The resulting spectrum is called $g(E_x)$ hereafter. 
Similarly, a second spectrum $g_>(E_x)$ is created by the convolution with a Gaussian function, whose width $\sigma_>$ is at least two times larger than the energy resolution in the experiment in order to remove gross structures from the spectrum.
It has been found that the most stable results are obtained by setting $\sigma \simeq 0.5\cdot \Delta E$ and $\sigma_{>} /\sigma = 2.5 -3.5$ in agreement with the results of Refs.~\cite{kal07,usm11b}.   
The spectra $g(E_x)$ and $g_>(E_x)$ for the present data are shown in Fig.~\ref{fig:dwtprocedure}(b).
The dimensionless stationary spectrum $d(E_x)$ defined by
\begin{equation}
    \label{eq:stsection}
        d\left( {E_x } \right) =
        \frac{{g_> \left( {E_x } \right)}}{{g \left( {E_x } \right)}}
\end{equation}
is shown in Fig.~\ref{fig:dwtprocedure}(c). 
As a result of the normalization on the local mean value, the energy dependence of the cross sections vanishes. 
The value of $d(E_x)$ is sensitive to the fine structure of the spectrum and distributed around an average intensity $\langle d(E_x)\rangle$=1. 
With increasing excitation energy the mean level spacing is decreasing, and in turn the oscillations of $d(E_x)$ are damped.
A quantitative description of the fluctuations is given by the autocorrelation function
\begin{equation}
    \label{eq:autocorexp}
        C\left( \epsilon  \right) =
        \frac{{\left\langle {d\left( {E_x } \right) \cdot d\left( {E_x  +
        \epsilon } \right)} \right\rangle }}{{\left\langle {d\left( {E_x }
        \right)} \right\rangle  \cdot \left\langle {d\left( {E_x  +
        \epsilon } \right)} \right\rangle
        }}\;.
\end{equation}
The value $C(\epsilon = 0) - 1$ is nothing but the variance of $d(E_x)$
\begin{equation}
    \label{eq:autocorvar}
        C\left( {\epsilon  = 0} \right) - 1 = \frac{{\left\langle {d^2
        \left( {E_x } \right)} \right\rangle  - \left\langle {d\left( {E_x
        } \right)} \right\rangle ^2 }}{{\left\langle {d\left( {E_x }
        \right)} \right\rangle ^2 }}\;.
\end{equation}
According to Ref.~\cite{jon76}, this experimental autocorrelation function shown in Fig.~\ref{fig:dwtprocedure}(d) can be approximated by an expression
\begin{equation}
\label{eq:autocorrtheo}
C(\epsilon) - 1 =  \frac{\alpha \cdot \langle \mbox{D} \rangle}{2 \sigma \sqrt{\pi}} \times f(\sigma,\sigma_>),
\end{equation}
%
%\begin{widetext}
%
%\begin{eqnarray}
%    \label{eq:autocorrtheo}
%        C(\epsilon) \; - \; 1 
%        \;\;\; = \;\;\; \frac{\alpha \cdot \langle \mbox{D} \rangle}{2
%       \sigma \sqrt{\pi}} \; \times & & \!\!\!\!\!\!\!\!\! \left\{ \exp
%        \left( -\frac{\epsilon^2}{4\sigma^2} \right) \; + \; \frac{1}{y}
%        \cdot \exp \left( -\frac{\epsilon^2}{4 \sigma^2 y^2} \right) \; -
%        \right. \nonumber \\ & & \nonumber \\ & & \!\!\!\!\!\!\!\!\!
%        \left. \sqrt{\frac{8}{1+y^{2}}} \cdot \exp \left( -
%        \frac{\epsilon^{2}}{4\sigma^2 (1+y^{2})} \right) \right\},
%\end{eqnarray}
%
%\end{widetext}
%
where the function $f$ depends on experimental parameters only.
The value $\alpha$ is the sum of the normalized variances of the assumed spacing and transition width distributions
\begin{equation}
    \label{eq:alpha}
        \alpha  = \alpha _{D}  + \alpha _{I}\;.
\end{equation}
If only transitions with the same quantum numbers $J^\pi$ contribute to the spectrum, then $\alpha$ can be directly determined as the sum of the variances of the Wigner and Porter-Thomas distributions, $\alpha = \alpha_{W} + \alpha_{PT} = 0.273 + 2.0$. 
The mean level spacing $\langle D\rangle$ is proportional to the variance of $d(E_x)$ and can be extracted from the value of $C(\epsilon = 0) - 1$. 
The nuclear level density can then be determined from the mean level spacing as $\rho(E)=1/\langle D\rangle$.

Uncertainties in the extracted values of $\left \langle D \right \rangle$ result from the following sources: (i) statistical errors, (ii) neglect of states of different $J^{\pi}$ or inaccuracy in the background determination, (iii)  widths of the smoothing functions, and (iv) finite length of the energy interval.
Statistical errors are negligible because of the folding of the spectra described above.  
Background contributions from other multipolarities to the $(p,p')$ cross sections at extreme forward angles are very small except for $M1$, and the $M1$ strength is confined to an excitation energies $E_{\rm x} \leq 9$ MeV outside the region analyzed here \cite{tam11,pol12}.
The determination of the background uses two independent techniques. 
Results for both methods are shown and their variation gives an estimate of the corresponding uncertainty.    
The choice of $\sigma$ and $\sigma_{>}$ gives rise to uncertainties in the mean level spacing of about $5\%$. 
The length of the interval is important, since too short intervals would result in errors in the autocorrelation function because of the finite number of data points. 
On the other hand, the exponential energy dependence of the level spacing within an interval is replaced by a linear one in the analysis, which is a reasonable approximation for sufficiently small ranges only.
The value of 0.5~MeV chosen in this study represents a compromise and limits the error contribution to about 2\%.

\subsection{Background determination}
\label{subsec:DWTBackground}

\begin{figure*}[tbh]
        \centering
\includegraphics[width=14cm]{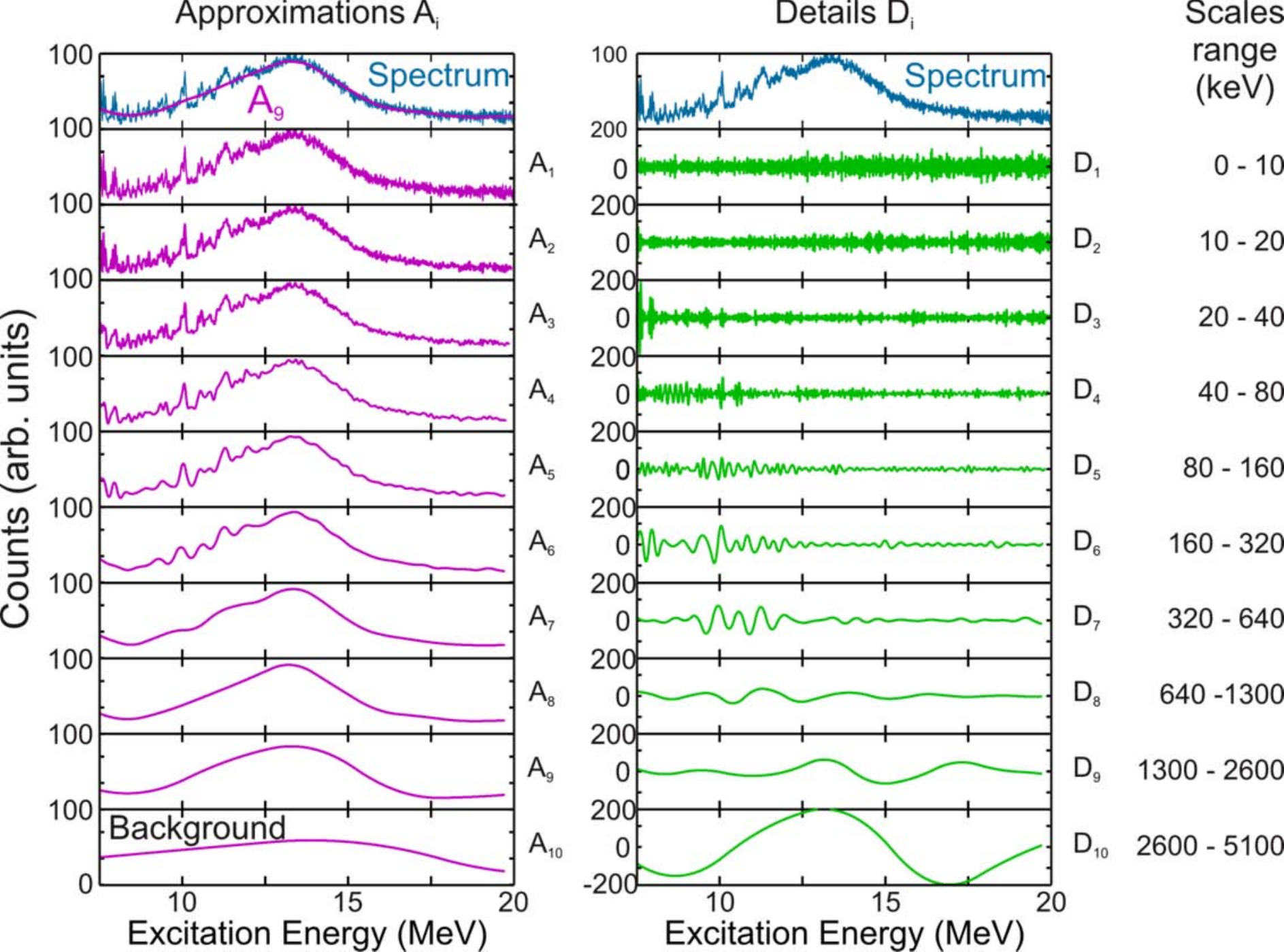}
        \caption[Decomposition of the $^{208}$Pb($p,p^\prime$) spectrum on approximations A$_i$ and details D$_i$]{(Color online) Decomposition of the $^{208}$Pb($p,p^\prime$) spectrum with the DWT analysis into approximations A$_i$ and details D$_i$. The approximation A$_9$ describes the total width of the GDR, thus A$_{10}$ can be adopted as background shape.}
                \label{fig:DWT_decompositionAD}
\end{figure*}
Two methods are applied to determine the background in the $(p,p')$ spectra due to nuclear processes. 
The first one uses the MDA described in Refs.~\cite{tam11,pol12}.
In the energy region of the IVGDR contributions other than $E1$ were found from excitation of the ISGQR and from a phenomenological background determined at excitation energies beyond the giant resonace region. 
Their sum represents the first background model. 
It is shown as dashed line in Fig.~\ref{fig:dwtprocedure}(a).
Alternatively, a spectrum decomposition based on the discrete wavelet transform (DWT) is used, where scales and positions in the wavelet analysis are varied by powers of two.   
It allows an iterative decomposition of the spectrum by filtering and decomposing it into two signals, approximations (A$_i$) and details (D$_i$). 
Application of the method to the spectrum of the $^{208}$Pb($p,p^\prime$) reaction is shown in Fig.~\ref{fig:DWT_decompositionAD}. 
The approximation is the large-scale or low-frequency component of the signal, and the detail corresponds to the small-scale or high-frequency part for a given scale region analogue to the effect of high- and low-pass filters in an electric circuit. 
In each step $i$ of the decomposition, the initial signal $\sigma$(E) can be reconstructed as
\beq
\sigma(E)=A_i + \sum D_i\;.
\label{eq:dwtapp+det}
\eeq
This operation can repeated until the individual detail consists of a single bin. 

A DWT can only be performed with wavelets which possess a so-called scaling function \cite{she08}.
This is not the case for the Morlet wavelet, thus the Bior3.9 wavelet function is used as an alternative.  
The BIOR wavelet family has a very similar form to the Morlet wavelet (cf.\ Fig.~9 in Ref.~\cite{she08}).
It also provides another useful property which can be applied for a determination of background in the data. 
Each wavelet function can be characterized by its number of vanishing moments,
\begin{equation}
   \label{eq:vanishingm}
   \int\limits_{ - \infty }^\infty  {E^n \Psi \left( E \right)dE =
   0,\;\; n = 0,1...m}.
\end{equation}
For Bior3.9 the number is equal to three, i.e., any background in the spectrum that can be approximated by a quadratic polynomial function does not contribute to the wavelet coefficients. 

\begin{figure*}[tbh]
        \centering
\includegraphics[width=14cm]{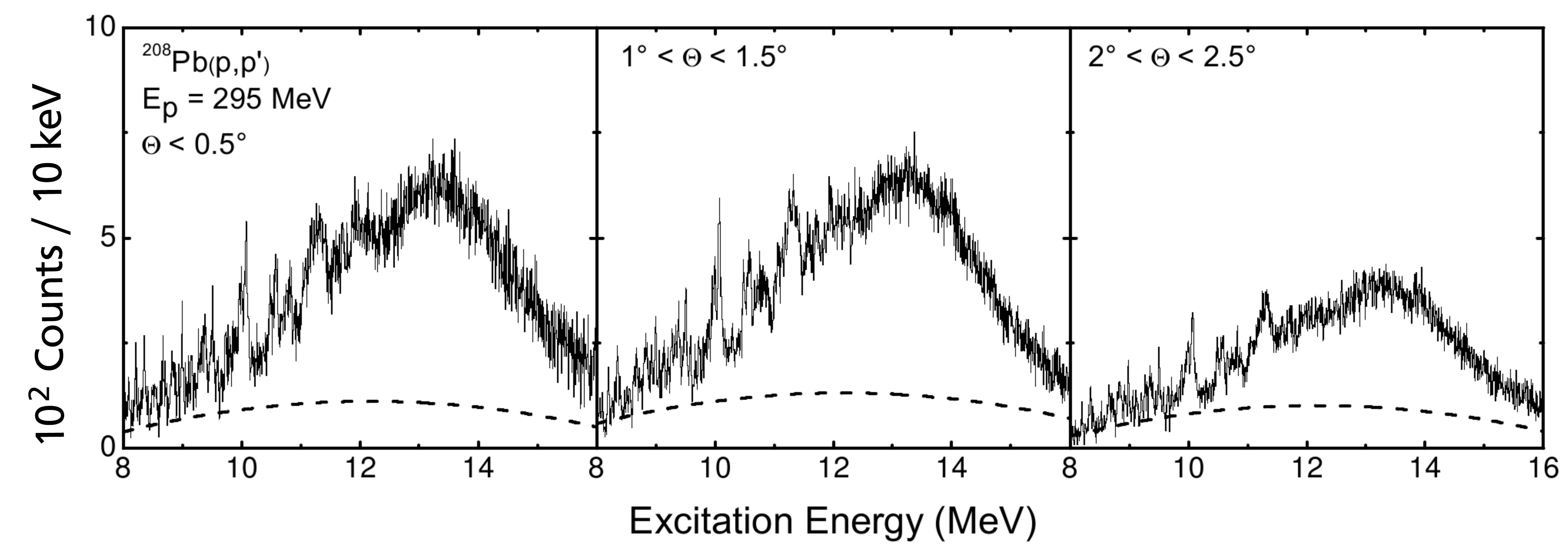}
        \caption[Excitation energy spectra with background from DWT]{Excitation energy spectra of the $^{208}$Pb$(p,p^\prime)$ reaction measured at the 0\dg~setting of the Grand Raiden spectrometer for different scattering angle cuts. The dashed lines show background exctracted by means of the DWT analysis.}
                \label{fig:DWT-bg}
\end{figure*}

The largest characteristic scale in the spectrum is given by the total width of the IVGDR.
It is well reproduced by approximation A$_9$ (cf.\ Fig.~\ref{fig:DWT_decompositionAD}).
Thus, the next approximation A$_{10}$ can be considered as a non-resonant contribution to the spectrum.
It determines the background except for an overall normalization, and the corresponding curve is shown in Fig.~\ref{fig:dwtprocedure}(a) as dotted line.
It is close to the background from the MDA analysis.
The normalization is determined from a repetition of the analysis for different angle bins.
Since the background shows a distinctively different angular dependence than the $E1$ cross sections, it can be fixed by the requirement of a constant level density in all spectra after background subtraction.
Figure~\ref{fig:DWT-bg} displays the resulting background shapes determined by means of the DWT analysis for three different scattering angle cuts obtained from the  measurement with the Grand Raiden spectrometer placed at $0^\circ$.

\subsection{Application to the excitation region of the IVGDR}
\label{subsec:leveldensityGDR}

\begin{figure}[tbh]
        \centering
\includegraphics[width=8.6cm]{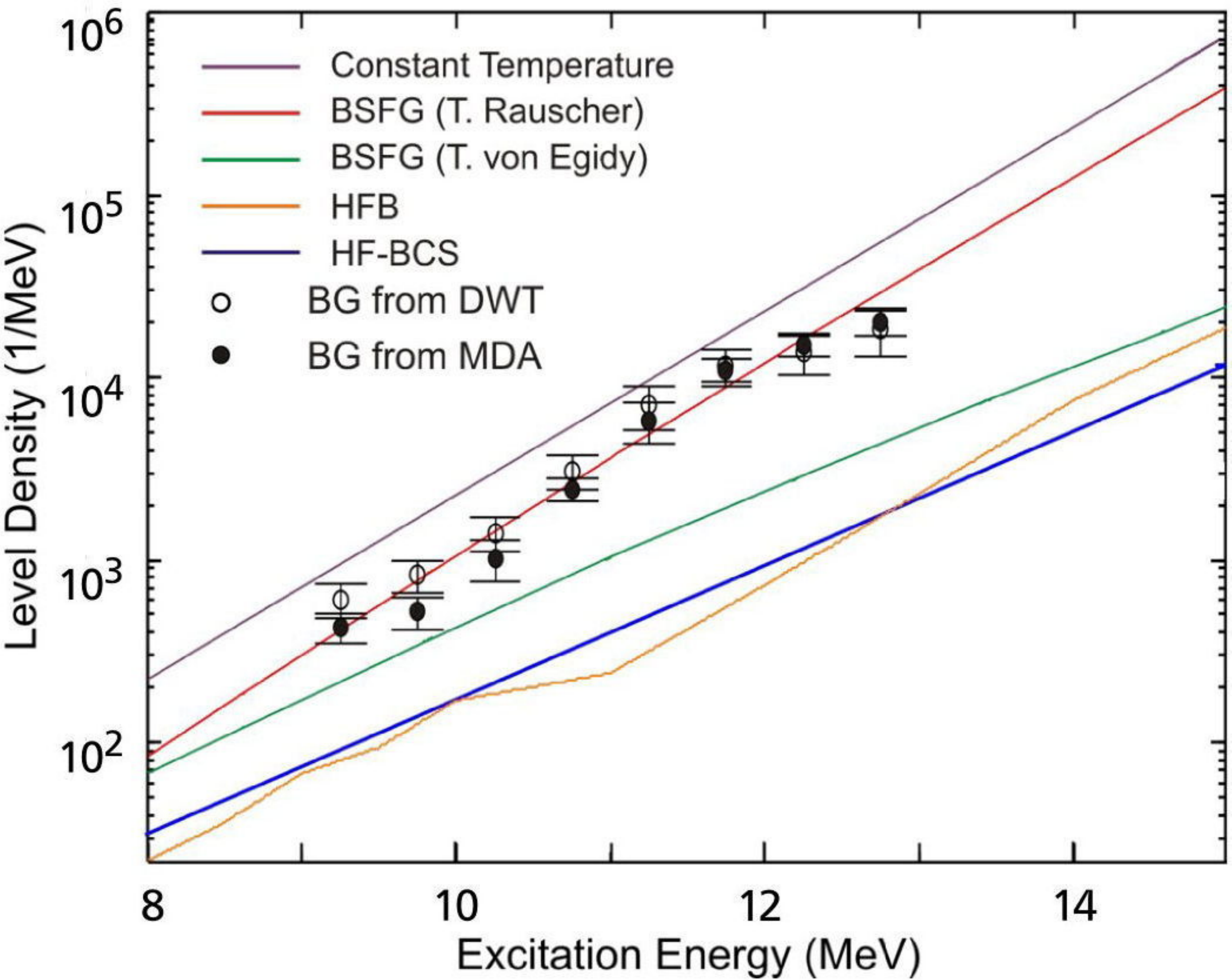}
        \caption{(Color online) Comparison of the experimentally obtained level densities for 1$^-$ states in $^{208}$Pb in the energy range from 9 to 12.5~MeV with predictions from BSFG using the models of~Ref.\cite{rau97} (red line)  and \cite{egi05} (green line), constant temperature model \cite{egi09} (purple line), HFB-BCS~\cite{dem01} (blue line) and HFB~\cite{gor08} (orange line).}
                \label{fig:LD}
\end{figure}

In Fig.~\ref{fig:LD} the experimental level densities of 1$^-$ states in $^{208}$Pb determined with the two different approaches of background subtraction are compared with values calculated by different phenomenological and microscopic approaches. 
For the fluctuation analysis the considered excitation energy interval between 8.5~MeV and 16~MeV has been split into subintervals of 0.5~MeV length.
The mean level spacing has been determined in each bin.
Insufficient statistics of the experimental spectrum or the onset of of the Ericson fluctuations in the excitation energy region above $~$12~MeV lead to a drop-down of experimentally deduced level densities. 
The phenomenon has also been observed in a similar analysis of $M2$ resonances in $180^\circ$ electron scattering data~\cite{vnc99}. 
Repetition of the analysis using different angle bins as described above suggests a comparable upper limit of the excitation energy in which the fluctuation analysis can be applied.
Therefore, the results shown in Fig.~\ref{fig:LD} are restricted to excitation energies $E_x = 9 -12.5$ MeV.

The experimentally obtained level densities are compared with different parametrizations of the phenomenological Back-Shifted Fermi Gas (BSFG) \cite{rau97,egi05} and constant temperature models \cite{egi09} and with microscopic calculations performed in the framework a Hartree-Fock-Bogoliuvov (HFB) \cite{gor08} or a Hartree-Fock-BCS approach~\cite{dem01}.
Good agrement with the BSFG parametrization of Ref.~\cite{rau97} is found.
The constant temperature model of Ref.~\cite{egi09} reproduces correctly the energy dependence but gives two times higher level densities.
All other models including the BSFG parametrization of Ref.~\cite{egi05} and the micoroscopic HFB \cite{gor08} and HF-BCS \cite{dem01} approaches fail to reproduce the magnitude and the energy dependence of the experimental data.

\section{Conclusions}
\label{sec:conclusions}

In the present work, the fine structure of the IVGDR in $^{208}$Pb observed nearly background-free in a high-resolution measurement of the $(p,p')$ reaction at $E_0 = 295$ MeV and $\theta = 0^\circ$ is investigated.
A wavelet analysis \cite{she08} reveals energy scales ranging from about 100 keV to several MeV characterizing the fine structure. 
Their nature and relation to dominant decay mechanisms can be interpreted by comparison to microscopic calculations of the $B(E1)$ strength distribution in $^{208}$Pb within the QPM \cite{rye02} and RTBA \cite{lit07} models.
For both approaches results including $1p1h$ and ($1p1h$+$2p2h$) model spaces are available. 
The fine structure and most prominent scales appear already at the RPA level.
The consistency of characteristic scales extracted with and without $2p2h$ states indicates that the coupling of the $1p1h$ doorway states to more complex states is weak and suggests Landau damping as the main source of fine structure of the IVGDR in $^{208}$Pb.
Since no additional scales appear with the inclusion of complex configurations, other mechanisms like direct decay or the coupling to low-lying collective vibrations identified as dominant mechanism inducing fine structure in ISGQR in heavy nuclei~\cite{she09} seem to play a minor role only.

The spectral fluctuations also provide information on the level density of $1^-$ states in the lower energy region of the IVGDR up to about 12.5 MeV.
Two different methods are applied for the subtraction of spectrum contributions not related to excitation of the IVGDR based on a discrete wavelet analysis and a multipole decomposition of the cross section angular distributions, respectively.
The results are not very sensitive to the particular choice since the background amounts to a few precent of the total cross sections only.
The consistency of the analysis is further demonstrated by the good agreement of level-density values obtained for different angular bins.
The fluctuation analysis method is applicable in the present case up to about 12.5 MeV; at higher excitation energies the statistics are insufficient and/or the Ericson regime of overlapping level widths is reached.
Still, a region of about 5 MeV above the neutron threshold can be covered complementary to most other methods restricted to energies below and close to threshold.   
The phenomenological BSFG model of Rauscher {\it et al.}~\cite{rau97} describes the experimental data well, while the BSFG approach of Ref.~\cite{egi05} and microscopic HF-BCS \cite{dem01} and HFB~\cite{gor08} give too low absolute values and also a weaker increase with excitation energy than experimentally observed.

The present study is another example of the power of high-resolution inelastic scattering studies of giant resonances.
The $(p,p')$ reaction at incident energies of a few 100 MeV and scattering angles close to $0^\circ$ is a remarkably selective tool for excitation of the IVGDR by relativistic Coulomb excitation \cite{tam09,nev11}.
The results presented above indicate a different mechanism leading to fine structure of the IVGDR than found in the ISGQR.
In the former Landau damping causes the pronounced structures observed in $^{208}$Pb between about 9 and 12 MeV.
For the latter resonance fine structure was shown \cite{she04,she09} to arise from the contribution to the spreading width due to coupling to low-energy surface vibrations \cite{ber83}.
For the case of the IVGDR it is important to study the fine structure systematically over a wider mass range in order to clarify to what extent the dominant role of Landau damping is a general phenomenon or related to the doubly closed-shell structure of $^{208}$Pb.
Work along these lines is underway. 
 
\begin{acknowledgements}
We are indebted to the RCNP for providing excellent beams. 
This work was supported by DFG (contracts SFB 634 and NE 679/3-1) and JSPS (Grant No.~14740154). 
E.L.\ acknowledges support by the US-NSF grant PHY-1204486 and National Superconducting Cyclotron Laboratory at Michigan State University.
\end{acknowledgements}

% Create the reference section using BibTeX:
%\bibliography{pbprl}

\end{document}